\documentclass[aps,prb,preprint,superscriptaddress,showpacs]{revtex4}
\usepackage{amsmath}
\usepackage{amssymb}
\usepackage{upgreek}
\usepackage{color}
\usepackage{graphicx}
\usepackage{subfig}
\newcommand{\fues}[1]{\left(#1\right)}
\newcommand{\bs}[1]{\mathbf{#1}}
\newcommand{\yav}[1]{\left[#1\right]}
\begin{document}

\title[]{Non linear magnetotransport theory and Hall induced resistance oscillations in  
graphene }

\author{Ricardo Guti\'errez-J\'auregui    }
\author{Manuel Torres}
\email[Email:]{torres@fisica.unam.mx} 
\affiliation{Instituto de F\'{\i}sica,
Universidad Nacional Aut\'onoma de M\'exico,
Apartado Postal 20-364,  M\'exico Distrito Federal 01000, M\'exico}

\date{\today}
\begin{abstract}

The quantum oscillations of nonlinear magnetoresistance in graphene that occur in response to a dc current bias are investigated. We present a theoretical model for the nonlinear magnetotransport of graphene carriers. The model  is based on the  exact solution of the effective Dirac equation in crossed electric and magnetic fields,  while the effects of randomly distributed impurities are perturbatively added. To compute the nonlinear current we develop a  covariant formulation of the  migration center theory. 
The analysis of the   differential resistivity in the large magnetic field region, shows that the extrema of the Shubnikov de Hass oscillations invert when the dc currents exceeds a threshold value.   These results  are   in good agreement with the experimental observations.  In the small  magnetic field regime, corresponding to large filling factors,  the existence of Hall induced resistance oscillations are predicted for ultra clean graphene samples. These oscillations  originate from  Landau-Zener tunneling  between  Landau levels, that are tilted by the strong electric Hall field.

\end{abstract}

\maketitle


\section{\small{ INTRODUCTION}}\label{intro}

Since the discovery of graphene\cite{novo:666},  a lot of excitement has been  generated, 
due to its unique electronic properties  and its potential applications in electronic devices\cite{kats:20,novo:1379}.
The dynamic of charge carriers in graphene is described by a  Dirac-like Hamiltonian, in the vicinity of the so-called Dirac points (where the conduction and valence band merge), leading to  an energy spectrum that has linear dependence on momentum  \cite{wall:622,seme:2449}.  The linear dispersion relation,   simulates the physics of quantum relativistic effects for chiral massless fermions, except for the fact that in graphene the Dirac fermions move with a speed $v_F$, which is 300 times smaller than the speed of light $c$. Hence, graphene exhibits a variety of pseudo relativistic phenomena, 
providing a  connection between condensed matter physics and  quantum-relativistic phenomena  \cite{castro:109}. Among others we can cite:  the   Zitterbewegung and its relation with  the minimal electrical conductivity at vanishing carrier concentration, the unconventional quantum Hall effect,  and the Klein tunneling  \cite{castro:109}.

The study of magnetic field effects on  the  electron gas dynamics has attracted the attention of physicists for long time. Magnetic oscillations,  such as the Shubnikov-de Haas (SdH) oscillations of the electric conductivity, and the de Haas-van Alphen oscillations of magnetization, are  a useful tool to study the shape of the Fermi surface \cite{ashmer}.  Furthermore, it has been found that magnetotransport phenomena are enhanced when the electron gas is confined to two dimensions, leading to one of the most remarkable phenomena in condensed matter: the  quantum Hall effect (QHE)  \cite{vonK}. More recently,  non-equilibrium and non-linear magnetotransport  phenomena  in high mobility two-dimensional electron gases (2DEG)  have acquired great experimental and theoretical interest. Microwave-induced resistance oscillations (MIRO) and zero resistance states  were discovered \cite{zudov2,mani1} in 2DEG  subjected to microwave irradiation and moderate magnetic fields. An analogous effect,  Hall field-induced resistance oscillations (HIRO) and zero differential resistance states,  
 have been observed in high mobility samples in response to a dc-current excitation \cite{Yang,Bykov}. 

Analogous phenomena to those described for conventional 2DEG in the previous paragraph, have been discovered in graphene,  including the integer  \cite{novo:197,zhan:201} and fractional  quantum Hall effect  \cite{kim:2009}.
However the non-linear magnetotransport effects in graphene are less studied.  
A recent experiment\cite{Tan} has analyzed the effect of a strong dc current bias on SdH  oscillations in graphene. 
A phase inversion on the SdH oscillations of the differential resistance is observed with increasing bias, in which the oscillation maxima evolve into minima and vice versa. On the work  of Wang and Lei\cite{wale} the nonlinear effects on graphene magnetotransport are studied within the balance-equation approach.

In this paper we analyze  a theoretical model for the nonlinear magnetotransport of graphene carriers. The model 
is based on the  exact solution of the effective Dirac equation in crossed electric and magnetic fields, 
while the effects of randomly distributed impurities are perturbatively added. We take advantage of the relativistic-like structure of the model to compute the nonlinear current. With this purpose  we develop a  covariant formulation of the migration center theory. Based on this model the quantum oscillations of nonlinear magnetoresistance in graphene that occur in response to a dc current bias are investigated.
The analysis of the   differential resistivity in the large magnetic field region shows,  in good agreement with the experimental observations  \cite{Tan},  that the extrema of the Shubnikov de Hass oscillations invert when the dc currents exceeds a threshold value.  
 At  small  magnetic field, the existence of Hall induced resistance oscillations are predicted for ultra clean graphene samples. These oscillations  originate from  Landau-Zener transitions  between  Landau levels, that are tilted by the strong electric Hall field.

The outline of the paper is presented as follows. In Sec. \ref{mod} the effective Dirac equation for charge carriers  in graphene subject to  crossed electric and magnetic fields, as  well as randomly distributed impurities is described. We present the  method, based in Lorentz-like  transformations, that allow us to solve the  problem. Section \ref{curr} describes the migration center theory, and the computation of the impurity assisted current.   The details of the current controlled scheme that lead to the  expression for  the longitudinal resistivity are discussed in Sec. \ref{ccsch}. In Sec. \ref{res}  we present  the main results and  discussion of the  nonlinear effects  in  SdH oscillations as well as the Hall induced resistance oscillations in graphene. 
The formulation of the    covariant  migration center theory is described  in the appendix.

\section{\small{MODEL}}\label{mod}

We consider the motion of  charge carriers  in an imperfect graphene sample,  in the presence of an in-plane electric field  and a  magnetic field  perpendicular to the graphene plane. The graphene defects are modeled by the  impurity scattering potential $V$.
The goal  of this work is to study the nonlinear  response to  a DC electrical 
current  density  ($J_{dc}$)  of graphene  placed in  a magnetic field. In a typical experimental configuration the electric field is not explicitly controlled, instead the longitudinal current is fixed,  $J_x= J_{dc}$, while the transverse current 
cancels, $J_y= 0$. Then, in general, the electric field has both longitudinal  ($ E_x$)  and transverse ($ E_y$) components, that have to be determined in terms of $J_{dc}$.
 The calculations are simplified if we perform a rotation by an angle $\phi =  Arctan(E_y/E_x)$ to 
a  $S_L^\prime$ frame with coordinates $\textbf{r}^\prime = (x^\prime, y^\prime)$:, hence,  the electric field points along the 
$y\prime$ axis: $\textbf{E}=\left(0,E \right)$. In  section  \ref{curr} the current density $\textbf{J}^\prime$ is calculated from the wave function solution that is obtained in the present section, an inverse rotation to the original  $S_L$  frame will  lead  to a set of two implicit equation for the unknown $E_x$ and $E_y$, that can be determined by a self-consistent iteration in terms of $J_{dc}$.

The dynamic of the problem in the  $S_L^\prime$ frame  is governed a Dirac  equation, that for convenience is written as 
\begin{align}\label{eqDirac}
&    v_F \,  \Pi_0 \Psi  =    \, \left[  v_F \,  \boldsymbol{\sigma} \cdot \boldsymbol{\Pi}   + V \right]  \Psi , 
\end{align}
where   $v_F$ is the Fermi velocity and   $\boldsymbol{\sigma} \equiv (\sigma_x, \sigma_y)  $ are  the Pauli matrices that act on  the two  component spinor  $\Psi $, 
 the pseudo-spin degrees of freedom are associated to  the two graphene sub-lattices.
In the previous equation $ \Pi_0 = p_0 -  \frac{e}{v_F}  \phi $  and  $\boldsymbol{\Pi} = \textbf{p} -    \frac{e}{c}   \textbf{A}$.
When necessary, we use a covariant notation with space-time $2+1$-vectors, $e.g$:  $x_\mu^\prime = \left( v_F t , \textbf{r}^\prime \right)$; $\mu=0,1,2.$ The momentum operator is 
$p_\mu = \left( i \hbar \partial /  \partial x_0^\prime,  - i \hbar    \vec{ \nabla }   \right)$, and $\Pi_\mu = p_\mu -  \frac{e}{c} A_\mu$. Notice that 
 the gauge potential  for constant electric and magnetic fields is given as:  
$A_\mu = \left(  - \frac{c} {v_F}E  y^\prime  , - B y^\prime, 0 \right)$. The Dirac matrices  are  selected as  $\gamma_{\nu} = \left(\sigma_{z} , i \sigma_{y} , -i\sigma_{x} \right)$\cite{Sharapov}.

The impurity  potential is decomposed  in terms of its Fourier
components
\begin{equation}\label{imppot}
V\fues{\bs{r}^\prime}= 
e^{- \delta \vert t \vert /\hbar } \, \sum_i^{N_i}\int\frac{d^2q}{\fues{2\pi}^2}V\fues{q}
\exp\yav{i\bs{q}\cdot\fues{\bs{r}^\prime-\bs{r}_i^\prime}} \, , 
\end{equation}
where  $\bs{r}_i^\prime$ represents the impurity  positions, that are chosen at random;  $N_i$ is the number
of impurities; and the $\delta$ parameter determines the rate at which the perturbation is turned on and off. The Fourier transform of the disorder  potential,  for screened Coulomb interactions, can be approximated by   \cite{NomMcd} $V\fues{q}=   2  \pi  e^2  /   \epsilon  \left(  q + 4  \alpha_g k_F \right)   $
 where  $k_F$ is the Fermi momentum,  $ \alpha_g = e^2  /  \epsilon \hbar v_F$ is the effective fine structure constant and $\epsilon$ is the  material  permittivity.  However, as far as we do not consider regions very close to the 
 Dirac point  \cite{NomMcd}, the short range scattering  approximation can be used, leading to  a constant value   $ V\fues{q} \approx V_0 =  \pi \hbar^2 v_F / 2 k_F$.

The pseudo-relativistic structure of the Dirac equation  was exploited by Lukose $et. \,al.$\cite{Lukose}  in order to obtain an exact solution for the  equation that represents a  graphene charge carrier subjected to both magnetic and electric fields. The method requires  to apply a  Lorentz-like transformation to an auxiliary  reference frame $S_A$ where the electric field vanishes. This is the case if the relative velocity of $S_A$  with respect to the original frame $S_L^\prime$ is selected as the drift velocity $ \textbf{v}_D = c  \textbf{E}  \times  \textbf{B} / B^2$. 
Once the solution in $S_{A}$ is known it is possible to boost back to  $S_{L}^\prime$ in order to obtain the appropriate eigenfunctions and  eigenvalues of the problem. This is possible as long as the condition $ c E <v_{F} B$ is fulfilled. 

We follow the procedure presented in  the work of Lukose $et. \,al.$\cite{Lukose} and extend the method  
  to  take into account  the impurity scattering effects.   In order to cancel the electric field effect,  a boost along the $x^\prime$ axis
is performed,   
   with a boost parameter $\theta$ given  by
\begin{equation}\label{beta}
\tanh \theta = \beta= \frac{ c \, E}{v_{F} B} \, .
\end{equation}
Under the  Lorentz  boost, coordinates transform as covariant three-vectors: $\tilde{x}_{\mu} = \Lambda^{\nu}_{\mu} x_{\nu}^\prime$; where the explicit components of $\Lambda_{\mu}^{\nu}$ can be read from the following relations:
\begin{equation}\label{boost1}
 \left( \begin{array}{c}
\tilde{x}_0\\
\tilde{x}_1  \\
\tilde{x}_2 \end{array} \right) = \left( \begin{array}{ccc}
\cosh \theta & \sinh \theta & 0  \\
\sinh \theta & \cosh \theta & 0\\
0 & 0 & 1 \end{array} \right) \left( \begin{array}{c}
x_{0}^\prime\\
x_{1}^\prime \\
x _2^\prime \end{array} \right)  \,. \end{equation}
 Whereas, the wave function transforms as a Dirac spinor
\begin{equation}\label{boost2}
\Psi\left(x^\prime \right) \rightarrow  \tilde{\Psi}\left(\tilde{x}  \right)  =   \exp\left[ - \frac{\theta}{2}\sigma_x \right] \Psi\left( x^\prime \right) \, .
\end{equation}
By re expressing the Dirac  equation (\ref{eqDirac}) in terms of $ \tilde{ \Psi } (\tilde{ x })$ with the aid of  Eqs. (\ref{boost2}) and (\ref{boost1}) one finds that $ \tilde{ \Psi } (\tilde{ x })$ satisfies an equation with the same structure 
 \begin{equation}\label{eqDirac2}
     v_F \,  \tilde{\Pi }_0 \Psi  =    \, \left[  v_F \,  \boldsymbol{\sigma} \cdot \tilde{\boldsymbol{\Pi}}   + \tilde{V} \right]  \Psi , 
\end{equation}
with $   \tilde{\Pi }_\mu =  \tilde{ p}_\mu -  \frac{e}{c}  \tilde{ A}_\mu $, but now the gauge potential is given as 
$ \tilde{A}_\mu = \left(0,  - \tilde{B}  \tilde{y} , 0  \right) $, where $\tilde{B} =  B  \, \sqrt{1 - \beta^2} $. Hence in the boosted  
 frame   the charge carrier dynamics corresponds to a problem with vanishing electric field and  a reduced magnetic field
$\tilde{B}  $. Whereas the transformed impurity potential takes the form
\begin{align}\label{eq:potimpsa}
\tilde{V} \fues{\bs{\tilde{r}},\tilde{t}}= {\cal M }\frac{1}{\gamma} e^{-\delta |\tilde{t}  | / \hbar \gamma} \sum_i^{N_i}\int   \frac{\text{d}{\text{q}}^{2}}{(2\pi)^2} V(q) e^{-i {\textbf{q}}\cdot (\tilde{\textbf{r}}-\tilde{\textbf{r}}_i)}\, 
  \exp\left[ -i \beta  v_{F} q_{x} \tilde{t} 
 \right]\,.
\end{align}

here $\gamma  \equiv \cosh\theta = 1 /   \sqrt{1 -   \beta^2} $,  and  the matrix  $ {\cal M } $ is given by 
\[ {\cal M } =  \gamma^{0}    e^{\left(\theta \sigma _x /2 \right)}  \gamma^{0}    e^{\left(- \theta \sigma _x /2 \right)}  = \gamma\left(  \textbf{1} + \beta \sigma_x   \right)  .  \]
Notice that in the $S_A$ frame  the impurity potential acquires an additional  time-dependent term; the boost induces a superposition of  oscillatory terms, each  with a momentum  dependent frequency
$\omega_{q} =   \beta v_{F} q_{x}$.

When  the impurity potential is not included, Eq. (\ref{eqDirac2}) is exactly solvable. The eigenstates and spectrum  are  given by the well-known relativistic Landau levels (LL): 
\begin{align}\label{eq:eigenfuncionesS}
&\tilde{\Psi}_{n, \tilde{k}_x}  = {e}^{i  \tilde{k}_x  \,  \tilde{x} } \left( \begin{array}{c}
 {\text{sgn}}(n)\phi_{|n|-1}\left( \tilde{y} -  \tilde{Y}_{k_x}  \right)\\
i\phi_{|n|}\left( \tilde{y}  -  \tilde{Y}_{k_x}   \right)\end{array} \right), \nonumber\\
&\tilde{ {\cal E} } _n \,  =  \, \text{sgn}(n) \,   \frac{ \hbar v_{F}}{l_{\beta}} \sqrt{ 2 \vert n \vert} \, ,
\end{align}
here the Landau number $n$  is an integer and  the  quantum number  $\tilde{k}_x $  corresponds to the translation symmetry along  the $x^\prime$ direction. The function $\phi_n $ are the displaced harmonic oscillator eigenfunctions, centered at $ \tilde{Y}_{k_x} =  l_{\beta}^{2} \tilde{k}_{x}$; and the magnetic length in the $S_A$ system, 
$ l_{\beta}$, is related to the magnetic length $  l_{c} =   \sqrt{ c\hbar /e B } $  by the relation $l_\beta = l_c  \gamma^{1/2}$. Notice that the energy spectrum is degenerate with respect to $\tilde{k}_x $. 
Applying the inverse boost to the  solution in Eq. (\ref{eq:eigenfuncionesS}),  yield the eigenfunctions and spectrum in the original $S_L^\prime$ frame:
\begin{align} \label{eq:eigenfunciones}
&\Psi_{n,k_{x}}(x^\prime,y^\prime) = \text{e}^{i k_x x^\prime} \text{e}^{ \frac{\theta}{2}\sigma_x} \left( \begin{array}{c}
\text{sgn}(n)\phi_{|n|-1}\left(y^\prime - Y_{n,k_x} \right)\\
i\phi_{|n|}\left(y^\prime - Y_{n,k_x} \right)\end{array} \right), \nonumber\\
&{\cal E}_{n,k_x} = \frac{\hbar v_{F}}{l_{c}}\text{sgn}(n)\sqrt{2 |n|}(1-\beta^{2})^{3/4}-\hbar v_{F} \beta k_x \, ,
\end{align}
notice that, as expected,  the electric field breaks the LL degeneracy, the Landau levels are  tilted with respect to $k_x$ with a slope given by $  \hbar v_{F} \beta$.
The displaced harmonic oscillator wave function is now centered at $Y_{n,k_x}$, given as 
\begin{align}
&  Y_{n,k_x}= \  l_{c}^2 k_x + \frac{\text{sgn}(n)\sqrt{2|n|} \beta l_c}{(1-\beta^{2})^{\frac{1}{4}}}  .\nonumber
\end{align}
 The  wave function  center position  depends  both on $k_x$ and on the LL number $n$,  as a result the LL are mixed, in particular the electric field mixes the particle and hole solutions, consequently the eigenfunctions are not orthogonal.

In order to obtain the current, within  the linear response formalism, it is convenient to use an orthogonal base,  otherwise the algebra turns rather cumbersome. To avoid this problem and since the eigenfunctions in $S_{A}$ are orthogonal, it is convenient to work out  the current density $\tilde{J}_\mu$  in the $S_{A}$ frame,  and then  apply a Lorentz transformation to obtain the current density  $J_\mu$ in the original  $S_{L}^\prime$ frame.    


\section{\small{ CURRENT DENSITY IN THE NONLINEAR  REGIME  }}\label{curr}

We now turn to the calculation of the current density  in the boosted  $S_A$ system where the dynamic is ruled by  Eq. (\ref{eqDirac2}). 
To consider the dynamic of an electron in a magnetic field it is  customary to use the migration center theory \cite{Kubo,Champel,Torresk}, where the electron 
 coordinate $\textbf{r} $ splits into the guiding center $\textbf{X}=(X,Y)$ and the   cyclotron or   relative coordinate  $ \boldsymbol{\eta}$, $i.e.$ $\textbf{r} = \textbf{X} +  \boldsymbol{\eta}$, where $ \boldsymbol{\eta}=(-\frac{c  \Pi_y}{e B}, \frac{c \Pi_x}{e B})$.  With this decomposition the  $\eta_x$ and  $\eta_y$   coordinates
become non-commutative, similarly for $X$ and $Y$ : 
\begin{equation} \label{conmrel}
\left[\eta_x, \eta_y  \right] = i \frac{c \, \hbar}{eB} , \hskip1.0cm \left[X, Y  \right] = - i \frac{c \, \hbar}{eB}, \hskip1.0cm  \left[\eta_i, X_j  \right]=0 \, . 
\end{equation}

It is possible to develop  a covariant extension of migration center  theory  using the Schwinger  proper time formalism. In what follows we present the general aspects 
of the  covariant  migration center  theory,  details are 
worked out   in the appendix. The  covariant expression for the relative coordinate   $ \eta_\mu $ is selected as 

\begin{equation} \label{eq:covrelative}
 \eta^{\mu}  = \frac{1}{e}\frac{ F^{\mu \nu}   \Pi_{\nu}  }{F \cdot F } \,,
\end{equation}
where  the electromagnetic tensor
$F_{\mu \nu}$ is   constructed from the constant $ \textbf{E}$ and  $\textbf{B}$ fields as given in Eq.(\ref{fmunu}), 
and the  quantity $F \cdot F \equiv F_{\mu \nu} \, F^{\mu \nu} = B^2 - \frac{c^2}{v_F^2} E^2    \equiv  \tilde{B}^2$ is an invariant under the Lorentz transformation.  In the  $S^\prime_L$ frame the $2+1$-vector $\eta^{\mu}$ is  given by 

\begin{equation} \label{eq:relativeSL}
 \eta^{\mu}  =   \frac{\gamma^2}{e}   \left(  - \beta \Pi_y \, \, , \, \,  \frac{c \Pi_y}{B} \, \, , \, \,  \frac{c \Pi_x}{B} +\beta \Pi_0  \right)  \,\,   \,,
\end{equation}
where we recall  that  the electric field  points along the $y-$axis and  $\beta = c E/ v_F B$. Instead in the  $S_A$ frame, the relative coordinate is  given by  $ \tilde{ \eta}^{\mu}   =  (c/e \tilde{B})(0, -\tilde{ \Pi}_y, \tilde{\Pi}_x)$. It is  easily demonstrated that the relative coordinate transform as  a covariant $2+1$-vector under the Lorentz  transformations Eqs. (\ref{boost1}):  $\tilde{\eta}_{\mu} =   \Lambda_{\mu}^{\nu}  \eta_{\nu}$.

The three-vector  guiding center coordinate  is then defined as $X_\mu = x_\mu - \eta_\mu$.  The  guiding center  velocity is obtained from the  Heisenberg equation of motion (see the appendix), in the boosted  $S_A$ frame it reads:

 \begin{equation} \label{3velocity}
\frac{ d \tilde {X}_\mu}{d \tau}  = \gamma_0  \left( v_F \, , \,  - \frac{c}{e \tilde{B}}  \,  \frac{\partial \tilde{V} }{\partial \tilde {y} }  \, , \,     \frac{c}{e \tilde{B}}  \,  \frac{\partial \tilde{V} }{\partial \tilde {x} } \right)   \,.
\end{equation}
It is a straightforward exercise to prove (see the appendix)   that  the current density probability  $\cal{J}_\mu $ defined in terms of the guiding center  velocity as  $  { \cal J}_\mu  = \Psi^\dagger \gamma_0 \left(  d \tilde {X}_\mu / d \tau \right) \Psi$,  
is also   a  Lorentz $2+1$-vector.     

We now turn to the calculation of the current density. The current density  $ J_\mu$ is  computed 
from  the impurity,  thermal, and time  average of the   guiding center  velocity 
\begin{equation} \label{currave}
J_\mu =\left[  \frac{e}{S \, T } \int_{-T}^{T} \, dt  \langle {Tr}\left[ \rho  \dot{X}_\mu  \right]  \rangle \right]_{T \to \infty}  \,
\end{equation}
where $S$ is the sample area. In the $S_A$ frame the density matrix $\rho$ satisfies the Von Neumann equation:
\begin{equation}\label{eq:VonNeumann}
i \hbar \frac{\partial}{\partial t}\rho = \left[   v_F  \boldsymbol{\sigma} \cdot \tilde{\boldsymbol{\Pi}}    + \tilde{V},\rho \right] \,.
\end{equation}
The Hamiltonian in Eq. (\ref{eqDirac2}) splits into the first   part  that is exactly solved, Eq.(\ref{eq:eigenfuncionesS}),   and the impurity potential $\tilde{V}$,  the latter is treated as a perturbation. The density matrix is also decomposed, perturbatively,  as $\rho = \rho_{0} + \Delta \rho$, where $\rho_{0}$ takes the form 
$\rho_{0} = \sum_n \int \text{dk}\hspace{1mm} f_{n}|n, \tilde{k}_x \rangle \langle n,\tilde{k}_x|\,,$ 
where  $f_{n}$ is the Fermi distribution function.  Since the eigenfunctions $|\nu\rangle=|n ,\tilde{k}_x \rangle$  are  orthogonal and complete, the matrix elements of  $\Delta \rho$ are readily obtained as: 
\begin{equation}
\langle m,\tilde{k}'_x | \Delta \rho |n,\tilde{k}_x \rangle = \frac{\left( f_{n} - f_{m}\right) \langle m, \tilde{k}'_x |\tilde{V} |n,\tilde{k}_x\rangle }{ \tilde{ {\cal E} }_{m} -\tilde{ {\cal E} }_{n} -  \hbar q_{x}v_{F} \beta - 
 i \frac{\delta  }{ \gamma }} \,. \nonumber  
\end{equation}
$\tilde{ {\cal E}_{n}}$ is the energy of the state $|n,\tilde{k}_x \rangle$  given in Eq. (\ref{eq:eigenfuncionesS}).
If we consider randomly distributed impurities, the terms that have linear dependence on the impurity  potential  vanish once we take the average over the impurities. In the auxiliary reference frame $S_A$ the current space components   get  contributions that arise solely  from impurity scattering, it  is worked out as:
\begin{align}\label{corriente} 
 \tilde{ \mathbf{J}}^{imp}  \left[  E \right]  = \frac{e c g_{k} \, \text{n}_{i}}{  \gamma^2 \tilde{B}} \sum_{n,m} \int & \frac{ {d}^{2}q}{2\pi}  \, \left(\textbf{e}_{3}\times \mathbf{q}\right) \, \left( f_n - f_m \right) \,  
  \, G_{nm}(q_x) \,  |V(q)|^{2}  |\Uptheta_{mn}(\zeta)|^{2}  
 \,,  \end{align}   
where { $\text{n}_{i}$ is the impurity density, the LL degeneracy per unit area is given by  $g_{k} = 1/ ( 2 \pi l_{c}^{2} ) $
 and   $\textbf{e}_{3}$ is an unitary vector in the direction of the magnetic field. 
The transition matrix element  $ |V(q)|^{2}  |\Uptheta_{mn}(\zeta)|^{2}  $ includes the function 
\begin{align}\label{Landaus}
\Uptheta_{mn} = \gamma\left[D_{|m|,|n|}+\text{sgn}(nm)D_{|m|-1,|n|-1} \right] 
 -\beta \gamma \left[\text{sgn}(m)D_{|m|-1,|n|}+\text{sgn}(n)D_{|m|,|n|-1} \right] \,, 
\end{align}
here  the matrix elements   of the displacement operator
 $D_{nm} $, are given by the following expression 

\begin{equation}\label{cohere}
D_{nm}(\zeta)  = \left( \frac{n!}{m!}\right)^{\frac{1}{2}} \zeta^{m-n} e^{-|\zeta|^{2}/2} L_{n}^{m-n}(|\zeta|^{2}) \,, 
\end{equation}
where $ \zeta = \frac{l_{\beta}}{\sqrt{2}}(q_{x} + i q_{y})$. Finally, the function $G_{nm}$ displays a Lorentzian distribution
\begin{equation}\label{eq:Tnm}
G_{nm}(q_x) \equiv \frac{   \delta / \gamma  }{\left(\tilde{ {\cal E} }_m -   \tilde{ {\cal E} }_n - \hbar v_{F} \beta q_{x}\right)^{2} + \left( \delta / \gamma    \right)^{2}} \,,
\end{equation}
 the effect of the quasiparticle scattering  is introduced by means of $\delta$ that produces the broadening of the Landau levels. } In the $\delta \to 0$ limit,  $G_{nm}$  embodies the energy conservation condition as discussed below.  Due to the elastic scattering, the impurities transfer a 
 momentum  $q_x = \tilde{k}_x - \tilde{k}'_x$ to the electrons.  Recalling that the center of the oscillatory wave function is given as $ \tilde{Y}=  l_\beta^2 \tilde{k}_x $, it follows that the  momentum transfer is  equivalent to a hopping or shifting of the guiding center  $  \Delta \tilde{Y}= \tilde{Y} - \tilde{Y}'=  l_\beta^2 q_x $. In the $ \delta \to 0$ limit Eq. (\ref{eq:Tnm}) gives the resonance condition  $ \tilde{ {\cal E} }_n =  \tilde{ {\cal E} }_m +  \hbar v_{F} \beta q_{x}$.  If interpreted in the $S_A$ system, the inter-Landau transitions are  induced by the time dependent impurity potential Eq. ( \ref{eq:potimpsa}) that oscillates with the frequency  $\omega_q =  v_{F} \beta q_{x}$.  Alternatively, the results can be interpreted according to the displacement mechanism:  using the momentum  transfer    $q_x = \tilde{k}_x - \tilde{k}'_x$ and noticing that 
${\cal{E}}_{n,k_x}= \gamma \left(\tilde{{\cal{E}}}_{n} - v_{F}\hbar \beta \tilde{k}_x\right)$, the resonance condition can be rewritten as  an energy conserving condition in the $S_L$ frame 
    $  {\cal E}_{n,k_x} =  {\cal E}_{m,k'_x}  $, and the transition can be interpreted as a Landau-Zener tunneling between LL, tilted by the electric field Eq. (\ref{eq:eigenfunciones})\cite{Yang}.

 When the Lorentzian distribution is considered in Eq. (\ref{corriente}) the summation over the $n$ and $m$ indices does not converge. While it is customary to set a cut-off energy to avoid the convergence issue \cite{Kim}, we substitute the Lorentzian distribution with a Gaussian distribution, leading to a natural cut-off:
\begin{equation}\label{gaussi}
 G_{nm}(q_x)   \approx \frac{\gamma}{ 2\sqrt{\pi}\delta }\exp \left[- \gamma^2 \left(\frac{\tilde{ {\cal E} }_{m} -\tilde{ {\cal E} }_{n} - q_{x}v_{F}\hbar \beta}{ \delta  }\right)^{2} \right]. \nonumber
\end{equation}

Finally, the thermal and  impurity average of the time component of the  current density in  Eq. (\ref{currave})  is easily computed utilizing  Eqs. (\ref{3velocity}) and (\ref{eq:eigenfuncionesS}), yielding $\tilde{J}_0 = e \tilde{n}_e v_F$, where $ \tilde{n}_e $ is the electron density in the boosted frame. 
 
\section{\small{ CURRENT CONTROLLED SCHEME }}\label{ccsch}

We are now ready to obtain the current density $\textbf{J}$ in the laboratory frame,  from the corresponding current density $\tilde{J}_\mu$  in the auxiliary frame. $\tilde{J}_\mu$ was calculated in the previous section: the time component is simply given as 
$\tilde{J}_0 = e \tilde{n}_e v_F$; whereas the corresponding space components are obtained by computing the expressions contained in Eq.(\ref{corriente}). It is easy to verify that $\tilde{J}^{imp}_x$ cancels because of the angular integrations;  hence, we only need to compute the component  $\tilde{J}^{imp}_y$. Applying the inverse Lorentz transformation the current $J^\prime_\mu$ in the $S_L$   frame is obtained as:

\begin{equation}
 J^\prime_\mu = \left( e n_e v_F  \, , \, e n_e \frac{E}{B}  \,,  \,   \tilde{J}^{imp}_y \right)   ,  
 \end{equation}
  where the electron density $ n_e $  is related to the electron density in the boosted frame   according to the relation $ n_e  =   \gamma  \,   \tilde{n}_e $.  Recalling that the  imposed dc-current points along the $x-$ axis, we need  to perform an additional rotation by an angle   $\phi =  Arctan(E_y/E_x)$ (see the beginning of Sec. \ref{mod}). Consequently these leads to the following conditions: 

 \begin{eqnarray}\label{corrientes2}
&& J_{x}  =  \, \, \, \, \, \,  e  \text{n}_e  \frac{E_y  }{ B} \,+ \,  \cos \phi   \, \, \tilde{J}_{y}^{imp} \left[ E \right]  \equiv J_{dc}  \, , 
 \nonumber\\
&& J_{y }   =    -  \, e  \text{n}_e  \frac{E_x  }{ B} \,+  \, \sin  \phi  \,\,   \tilde{J}_{y}^{imp}  \left[ E \right]  \equiv 0 \,.
\end{eqnarray}
The last equalities apply for the Hall configuration:  the longitudinal current is fixed at the value $J_x= J_{dc}$, while the transverse current  cancels. Eqs. (\ref{corrientes2}) represent two implicit equations for the unknown 
$E_x$ and $E_y$. The  equations can be solved following a self-consistent iteration method. However, it is easily verified that for a  typical Hall experiment a correct  approximate   solution to the previous equation is obtained assuming that the electric field is almost transverse with respect to the dc bias current: $i.e.$ $E_x \ll E_y$ or $\phi \approx \pi/2 $. Retaining  terms to first order in $\epsilon = \pi/2 - \phi $, the explicit solution  is worked out as 

\begin{eqnarray}\label{solcorr}
&& E_{y} \approx E_H=  \frac{B}{e \text{n}_{e}} J_{dc} \, , 
 \nonumber\\
&& E_x     \approx    \frac{ B }{ e \, \text{n}_e  }  \, \,  \tilde{J}_{y}^{imp} \left[  \frac{B}{e \,  \text{n}_{e}} J_{dc} \right]  \,.
\end{eqnarray}
The first equality is obtained from  the leading contribution, it  sheds the well known result for the Hall electric field $E_H$ induced by the dc bias in the magnetic field.   The second equation follows from the next order contribution,  it gives   the longitudinal electric field in terms of the impurity scattering current obtained in Eq. (\ref{corriente}); the result allow us to calculate the longitudinal resistivity, $\rho_{xx} = E_{x}/ J_{dc} $, as well as the  longitudinal resistivity differential $ r_{xx} = \partial  E_{x}/ \partial  J_{dc} $.
\begin{equation}
\rho_{xx} = \frac{E_{x}}{ J_{dc}} \, ,    \hskip1.5cm r_{xx} = \frac{ \partial  E_{x}}{ \partial  J_{dc} } \,. 
\end{equation}

It is interesting to notice that the $\beta$ parameter, Eq. (\ref{beta}), takes the form:
$ \beta_H=  J_{dc} / e \text{n}_{e} v_{F} \,,$ 
so $\beta_H$ can be understood as the rate between the current density and charge carriers density moving at a top speed $v_{F}$. Clearly, the formalism  breaks down when $J_{dc}>e \text{n}_{e} v_{F}$.

\section{\small{RESULTS AND DISCUSSION}}\label{res}

Results are presented for  the differential resistivity $r_{xx}$ both in the strong and weak  magnetic field regimes. As explained in  section  \ref{curr}, the displacement mechanism is the  physical origin of the nonlinear effects: impurity scattering induces a spatial hopping of the charge carriers between tilted Landau levels. 
 However, there is an additional  effect that has to be consider.  Recent experimental results \cite{Tan} suggest that in graphene a current dc bias produces a strong electron  heating effect. It was found that the  charge carriers effective temperature depends linearly on the bias current as $T \sim   T_0  + \alpha J_{dc}$. The parameter $\alpha$ was estimated as $\alpha \approx 3 \ K \, / ( A/m)$, for bias currents $ 2 \sim 20  \,  A/m$. For stronger bias currents the relation is no longer linear \cite{Baker}.     
 
\begin{figure}[htbp]
  \begin{center}
\includegraphics[scale=0.5]{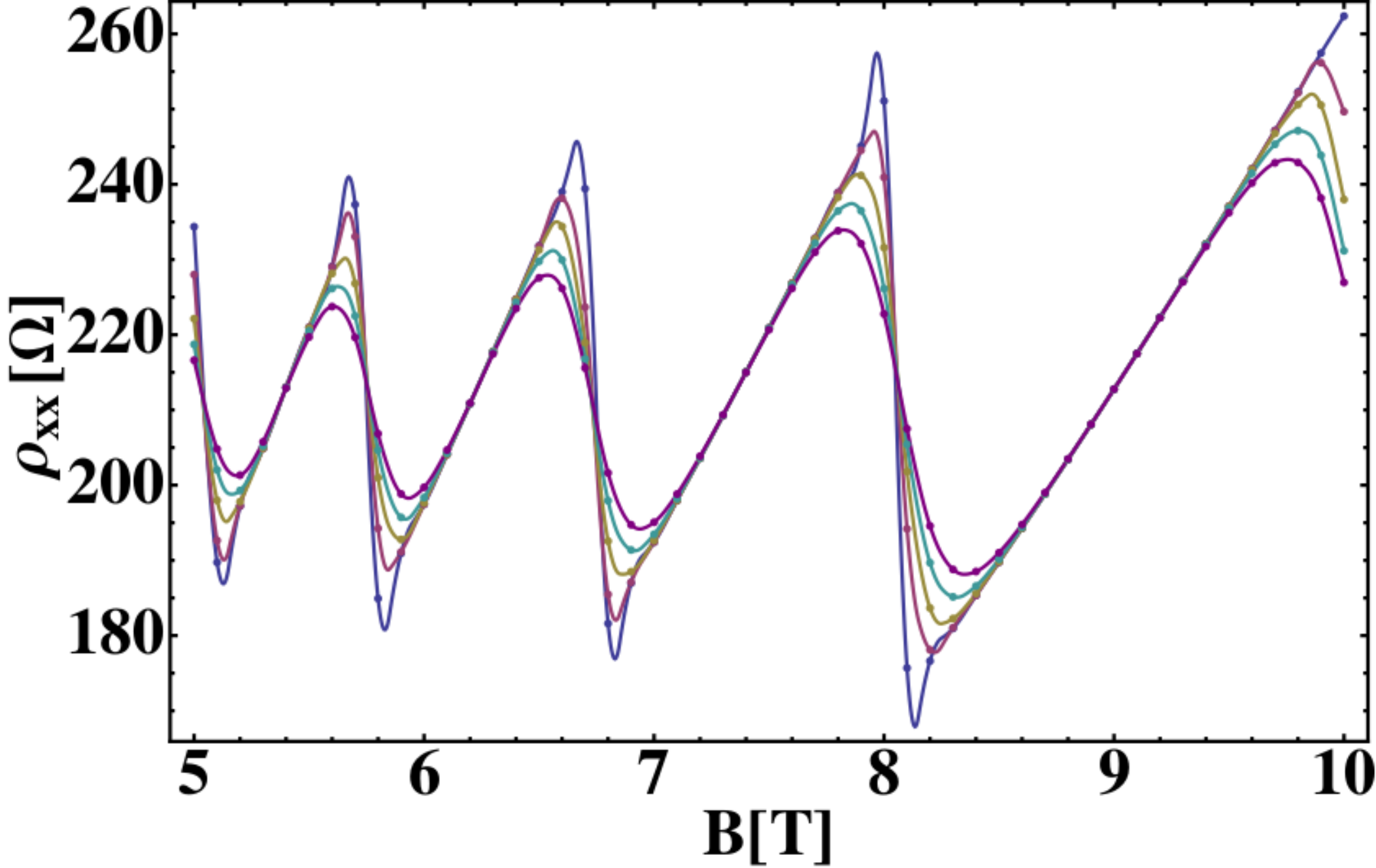}  
  \caption{ (Color on line)  Longitudinal resistivity  as a function of the magnetic field  at various  values of the dc current density, from top to bottom: $J_{dc} =  $ 
    $1.8 A/m$ (Blue), $ 3  A/m$ (Red),   $4.8  A/m$ (Yellow),  $5.5 A/m$(Green),  $6.7 A/m$ (Purple).  The other parameters are:   $n_e = 3   \times 10^{12} cm^{-2}$ and   $\tilde{\delta}=  0.14$. }
  \label{gresistencia}
  \end{center}
  \end{figure}

\begin{figure}[htbp]
  \begin{center}
\includegraphics[scale=0.5]{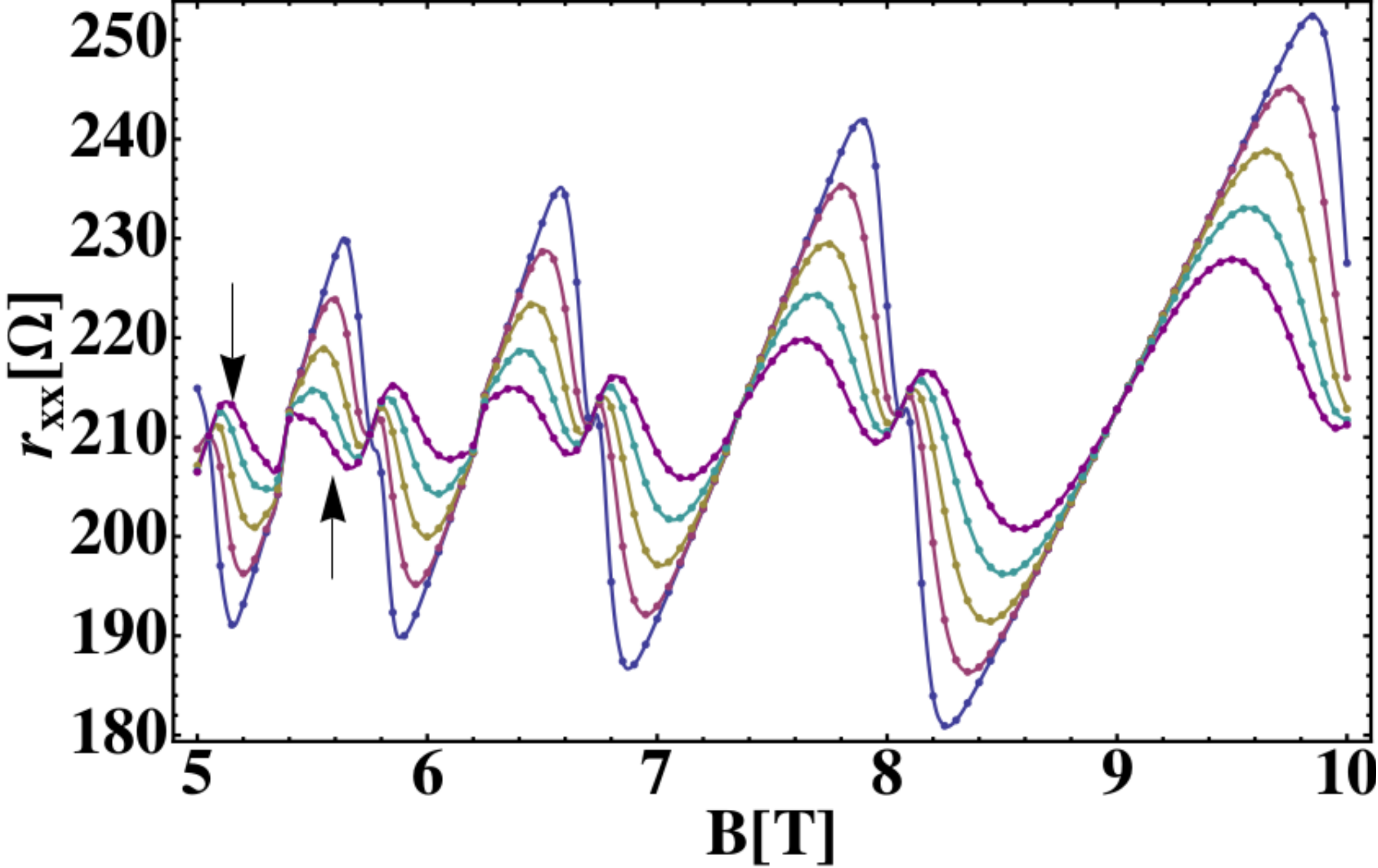}
  \caption{  (Color on line) Longitudinal  differential  resistivity, as a function of the magnetic field. Same  parameter values as in  Fig.  \ref{gresistencia}. From  top to bottom:   $J_{dc} =  $
    $1.8 A/m$ (Blue), $ 3  A/m$ (Red),   $4.8  A/m$ (Yellow),  $5.5 A/m$(Green),  $6.7 A/m$ (Purple). Observe the inversion of the  SdH oscillations extrema   in $r_{xx}$  as the current density is increased.}
  \label{gresistenciadiff}
  \end{center}
  \end{figure}

  First, we consider the nonlinear effects on the Shubnikov-de-Hass (SdH) oscillations  in the strong magnetic field region: $ B \sim 5-10 \, T$. The parameters are selected corresponding  to typical experiments  in graphene samples: $v_F= 1 \times 10^{6} m/s$, electron density $n_e = 3   \times 10^{12} cm^{-2}$,  and a  base temperature  $T_0 =2  \, K$.
 The width of the LL depends on the  broadening parameter $  \delta$, that  is calculated from a self-consistent Born approximation  calculations as  \cite{Ando,Krstajic}
 \begin{equation}{\label{delta}}
 \delta = \tilde{\delta}  \,  \sqrt{2}  \,   \frac{ \hbar v_F}{  l_c }   \,.
\end{equation}
 In a typical graphene sample with a mobility $\mu \approx 10^{4}  cm^{2} V^{-1} s{-1} $, the dimensionless parameter is estimated as $   \tilde{\delta} \approx 0.14$. 
 In Fig.  (\ref{gresistencia})   we present the longitudinal resistivity  $\rho_{xx}  (B)$ for various values of the  dc current $J_{dc}$. As observed in the experiments \cite{Tan}, the  amplitude of the SdH oscillations decreases with increasing temperature.  Fig.  (\ref{gresistenciadiff}) displays  the SdH oscillation for the differential resistivity as function of $B$ and for the same values of  $J_{dc}$. Besides the damping effect on the oscillations amplitude, it is observed that as $J_{dc}$ increases  the maxima of the oscillation evolve into minima, and  vice versa. These nonlinear effects on  SdH oscillations  originate mainly  from the current heating effect. If we compare the results of $r_{xx} (J_{dc})$ without the heating effect ($ \alpha =0)$,  with the result of the linear response approximation  $r_{xx} (0)$ we find essentially the same results, confirming the fact that the  displacement mechanism has a negligible effect on the  SdH oscillations.

\begin{figure}[htbp]
\begin{center}
\includegraphics[scale=1.0]{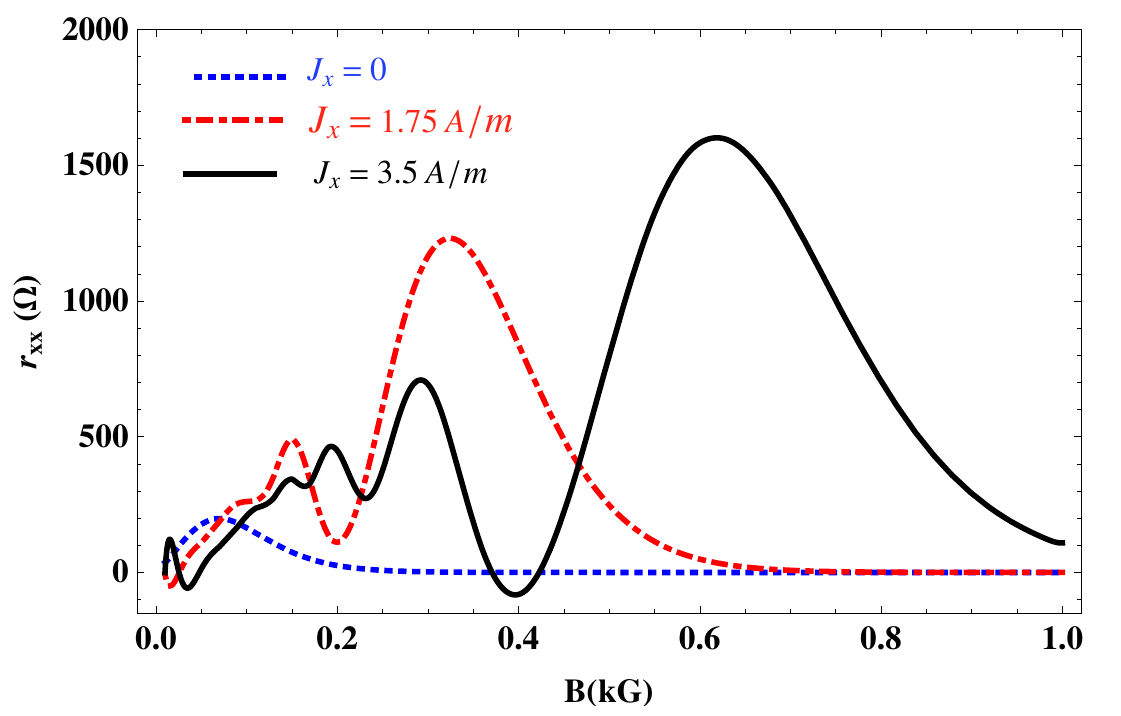}
\caption{Dependence of the differential resistivity on the  magnetic field in a high mobility graphene sample.   With no dc bias we observe the small decreasing curve  (dotted-blue). The two other curves correspond to  $J_{dc} =  1.75 \, A/m $ (dotted-dashed-red) and  $J_{dc} =  3.5 \, A/m $ (continuous-black). 
The other parameters are:   $n_e = 1   \times 10^{11} cm^{-2}$ and   $\tilde{\delta}=  0.014$.  }
\label{HIRO-1}
\end{center}
\end{figure}

 We now explore the possibility of detecting nonlinear oscillations in the low magnetic region. The study of ultra clean semiconductor samples 2DEG has shown that dc current excitation leads to Hall field-induced resistance oscillations \cite{Yang,Bykov,Kunold,Wiedmann}. In order to simulate a high mobility  graphene sample the value of  the LL width in Eq. (\ref{delta}) is reduced by an order of magnitude, by selecting    $   \tilde{\delta} \approx 0.014$; which would represent a sample with a high mobility $\mu \approx 10^{5}  cm^{2} V^{-1} s{-1} $.  Furthermore we consider an smaller value of the electron density $n_e = 1   \times 10^{11} cm^{-2}$.
 Fig.  (\ref{HIRO-1}) displays the differential 
 resistivity $r_{xx}$ as a function of $B$. First we notice the results of the linear response solution, that produces a small monotonous decreasing curve. However,  the two other curves obtained for  $  J_{dc}  = 1.75 A/m$ and  $  J_{dc}  = 3.5 A/m$,  display strong differential magnetoresistance oscillations. The physical origin of the intense HIRO  lies on the displacement mechanism, as explained below.

\begin{figure}[htbp]
\begin{center}
\includegraphics[scale=1.0]{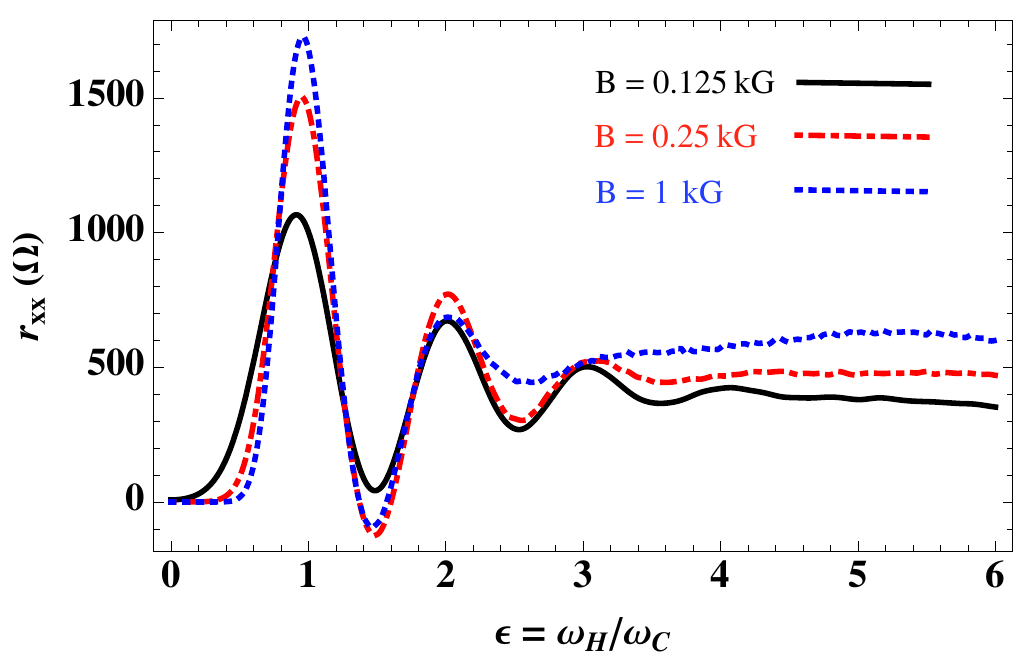}
\caption{Differential resistivity as a function of $\epsilon =  \omega_H / \omega_c $,
obtained from $J_{dc}$ sweeps ($J_{dc} \in [0.1, 6] \, A/m $) at fixed values of  $B =0.125 \, kG $, $B =0.25 \, kG $, and $B =1\,  kG $.
The other parameters are:   $n_e =  1  \times 10^{11} cm^{-2}$ and   $\tilde{\delta}=  0.014$.  }
\label{HIRO-2}
\end{center}
\end{figure}

The oscillations are periodic in $1/B$ and are governed by the ratio $  \epsilon =  \omega_H / \omega_c$ where the 
expression for the Hall frequency $\omega_H$ and a simple  explanation of the oscillation pattern follows from the  
  analysis of the structure of  $ \tilde{J}_{y}^{imp} \left( E_{H} \right) $ in Eq. (\ref{corriente}). For this purpose we consider the  limit $\delta \rightarrow 0$, in Eq. (\ref{eq:Tnm})  or Eq. (\ref{gaussi}), that leads to  the   resonance condition 
\begin{equation}\label{eq:Consenergia}
\frac{\Delta { \cal E }}{\hbar v_{F}} = \frac{\sqrt{2}  }{l_{\beta}}\left(\sqrt{n} - \sqrt{m}\right)- \beta \Delta k = 0 \,,
\end{equation}
  here  we are considering levels well above the Dirac point, so all the Landau numbers $n$ and $m$ are positive.  The  momentum transfer $ \Delta k$ is  equivalent to a hopping or shifting of the guiding center  $  \Delta Y= Y - Y'=  l_\beta^2 q_x $.    An estimation of the momentum transfer $\Delta k$ can be obtained   from the overlap of the wave   functions  between the $n$ and $m$ LL. The overlap, obtained in  section \ref{curr}   is proportional to the function $D_{nm}(\zeta )in $ Eq. (\ref{cohere}), 
where $\zeta  = \frac{l_{\beta}}{\sqrt{2}}(q_x + i q_y)$. While this functions is highly oscillatory, the envelope of the function  can be approximated by \cite{Slater,Zener}
\begin{equation}
\sqrt{\frac{n!}{m!}} \zeta^{m-n} e^{-|\zeta|^{2}/2} L_{n}^{m-n}(|\zeta|^{2}) \rightarrow  \frac{\zeta^{m+n} e^{-|\zeta|^{2}/2}}{\sqrt{n! + m!}} \, ,\nonumber  
\end{equation}
with a  maxima at $\zeta_{max} = \sqrt{m + n}$. Hence, the maximum overlap  corresponds to a momentum transfer of
$ \Delta k =  \sqrt{2(m + n)} /l_\beta$.
 Inserting this result  in Eq. (\ref{eq:Consenergia}) the following selection rule is obtained
\begin{equation}\label{eq:transgeneral}
\beta \left( \sqrt{n+m} \right) \left(\sqrt{n}+\sqrt{m} \right) = n-m .\,
\end{equation}
 In HIRO we are interested in large LL; so we consider contributions  from states that are close to the  Fermi level, that is far away from the Dirac point, so  $n_{F} \approx   \pi l_{c}^{2} \text{n}_{e}/ 2.$
 As we assume that   $n_{F} \gg 1$ we write   $m=n_{F}+ m^{\prime}$ and $n=n_{F}+ n^{\prime}$, with 
  $(m^{\prime},n^{\prime}) \ll n_{F} $.
  So the resonance condition (\ref{eq:transgeneral}) becomes
\begin{equation}\label{eq:transunomas}
\sqrt{2} \ \beta  \  \pi \,   \text{n}_{e} \ l_{c}^{2} \simeq N ,\,
\end{equation}
with $N =m^{\prime} - n^{\prime} $. 
We recall that, even though the linear spectrum in  graphene implies zero rest  mass, their cyclotron mass is not zero, but it is given as\cite{novo:197}  $m^{*}=\frac{\hbar \sqrt{\pi \text{n}_{e}}}{v_{F}}$. Hence,  
using the definition of the cyclotron frequency, $\omega_{c} \equiv \frac{e B}{m^{*}c}$,   the resonance condition can be written as
\begin{equation}\label{eq:resohall}
\sqrt{\frac{2\pi}{\text{n}_e}}\frac{J_{dc}}{e}  \equiv  \omega_{H} = N \omega_{c} \,.
\end{equation}  
The HIRO oscillations are  periodic in $1/B$, according to  the condition (\ref{eq:resohall}) resistance maxima are expected at  integer values of the dimensionless parameter   $  \epsilon =  \omega_H / \omega_c$. The Hall frequency 
$\omega_{H}$ can be interpreted in terms of   the energy $\hbar \omega_{H} = e  2 R_c  E_H$  
associated with the Hall voltage drop across the cyclotron diameter:  $E_H$ is  the Hall field as given in Eq.  (\ref{solcorr}) and  $R_c$ is the cyclotron radius at the Fermi level. To confirm these observations, in Fig.  (\ref{HIRO-2}) the differential resistivity is plotted as a function of $\epsilon$. Magnetoresistance oscillations are clearly observed up to the fourth resonance. The positions of the peak maxima appear in good agreement with  the  conditions in  (\ref{eq:resohall}).

 It was mentioned  in section (\ref{curr})  that according to the interpretation in  the $S_A$ system, the inter LL transitions are induced by the time dependent potential Eq. (\ref{eq:potimpsa}) that oscillates with the frequency $  \omega_q = v_F   \beta q_x$.  The transition matrix element peaks at the momentum transfer 
$ q_x = q^* = 2  \sqrt{n_F} /l_c $,  hence we find that the leading oscillation frequency of the periodic potential, is precisely  the Hall frequency $  \omega_{q^* } \equiv   \omega_H$, given in Eq.  (\ref{eq:resohall}).

It is also  interesting to observe that from Eq. (\ref{eq:Consenergia}) the hopping distance for a  resonant transition between LL is expressed as 

\begin{equation}
\Delta Y  =  \frac{  \sqrt{2} \gamma e n_e  v_F l_c }{J_{dc}}  \left(\sqrt{n} - \sqrt{m}\right)   \,.
\end{equation}
that depends both on the magnetic field and $J_{dc}$. This equation can be interpreted as a selection rule  that determines $\Delta Y $ for  arbitrary  transition between LL. If we now consider the limit of large LL levels, the selection rule reduces to  
\begin{equation}\label{eq:hoppdist}
\Delta Y  =  \frac{   \gamma e  \sqrt{n_e}  v_F }{\sqrt{\pi}J_{dc}}  \, N   \,.
\end{equation}
 which does not depend on $B$. This result is similar to the   selection rule 
$ \Delta Y  =  N \,    (e \hbar  n_e )/ (m^* J_{dc})   $ found by  Yang $et \,  al.$  \cite{Yang} in their study of 
   HIRO in 2DEG.

\begin{figure}[htbp]
\begin{center}
\includegraphics[scale=1.0]{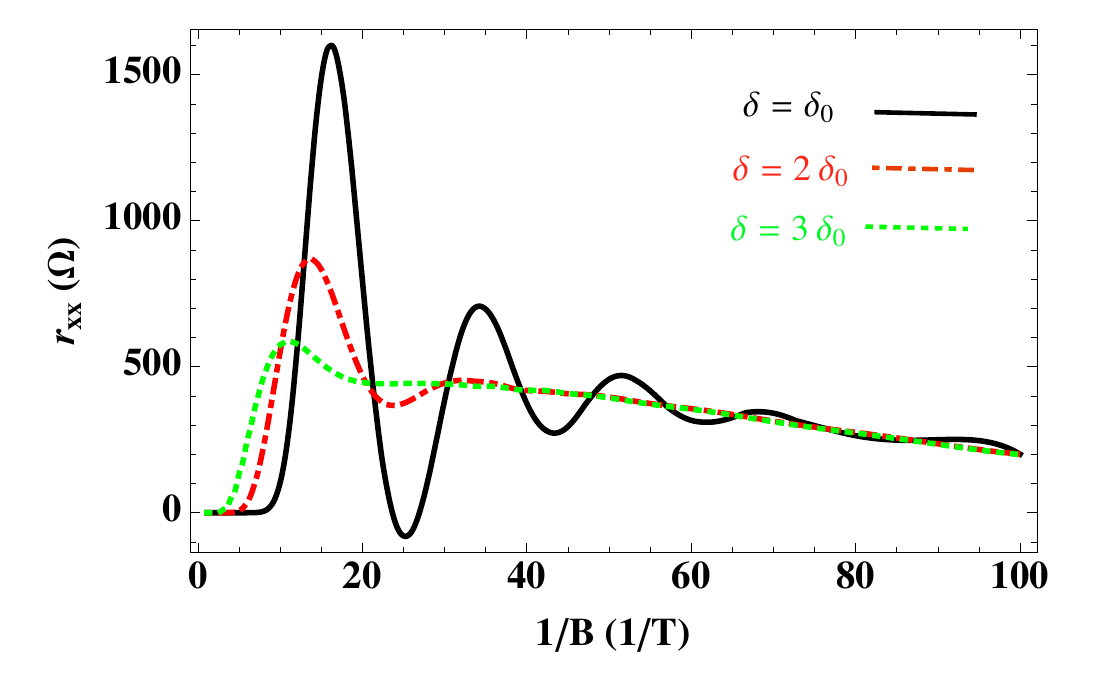}
\caption{Differential resistivity as a function of $1 / B$ for three values of the dimensionless broadening parameter in Eq. (\ref{delta}): 
 $ \tilde{\delta}  =0.014 $ (continuous, black),   $  \tilde{\delta}  =0.028 $  (dashed, red), and  $  \tilde{\delta}  =0.042$ (dotted, green). The other parameters are:   $n_e =  1   \times 10^{11} cm^{-2}$ and    $J_{dc}= 3.5 \, A/m $. }
\label{HIRO-3}
\end{center}
\end{figure}

To acquire a deeper insight on the experimental parameters required to obtain the HIRO,  in Fig.  (\ref{HIRO-3}) the differential resistance is plotted as a function of $1/B$  for  three values  of the dimensionless broadening parameters 
 $ \tilde{\delta}_1   = 0.014 $,   $ \tilde{\delta}_2   = 0.028 $, and   $ \tilde{\delta}_3   = 0.042 $, that 
  corresponds to a reduction of  the mobility by a factor of 2 and 3. Clearly, the HIRO amplitude are strongly  suppressed as $\delta$ increases. Although in the first case  the  HIRO are clearly identified, an small reduction of the mobility sweeps most of the oscillation.  Hence a well defined mobility threshold is expected. Producing ultra clean graphene samples with a mobility that exceeds the threshold value would probably lead to the experimental observation of HIRO.

\section{\small{CONCLUSION}}\label{conc}

We have presented a theoretical model for the nonlinear magnetotransport of graphene carriers. The response to a dc current bias is incorporated by the exact solution of the effective Dirac equation in crossed electric and magnetic fields, obtained by  means of a Lorentz transformation to a system  $S_A$ in which the electric field vanishes \cite{Lukose}. We  extend the method so the impurity scattering effects are taken into account, by a perturbative solution of the  density matrix equation. The nonlinear current is computed by means of the  migration center theory. The calculation is  carried out in the $S_A$ system,  and is  later  transformed to the original $S_L$ frame. For this purpose we develop a  covariant formulation of the migration center  theory.

The analytical expression for the impurity assisted current  Eq. (\ref{corriente}),  allow us to calculate the longitudinal  resistivity  and  the longitudinal differential resistivity. Based on these results the quantum oscillations of nonlinear magnetoresistance that occur in response to a current bias are investigated. When applied to the analysis of the  Shubnikov de Hass oscillations, it is observed that  the strong $J_{dc}$ produces a decay of the oscillations in the longitudinal resistivity Fig. \ref{gresistencia}.  Whereas  the differential resistivity plot, Fig. \ref{gresistenciadiff}, shows a phase inversion in which the maxima of the oscillation evolve into minima. Both effects are in good agreement with the experimental observations \cite{Tan}. It is demonstrated that  these  effects on  SdH oscillations  arise  from the current heating effect, while the contribution from the  displacement mechanism is negligible.

At the small  magnetic field region, corresponding to large filling factors, magnetoresistance  oscillation pattern  emerge. Analogous to what has been observed in 2DEG, the magnetoresistance oscillations are  induced by the intense Hall electric field (HIRO). 
The nonlinear effects produce a strong magnification of the differential resistance, as well as an oscillatory 
behavior that is  periodic in $1/B$. The oscillation are governed by the ratio  $\omega_c/\omega_H$.
The origin of the HIRO lies on the displacement mechanism.
The expression for the longitudinal current  in Eq. (\ref{corriente}) contains the main ingredients that explain the  displacement mechanism.  There are  two possible interpretations, depending on which frame it  is visualized: (i)  in $S_A$:   the relevant  LL  transitions are induced by the time dependent impurity potential, that oscillates with a distribution of frequencies $ \omega_q$ that peaks at the Hall frequency.
 Alternatively  (ii) in  $S_L$: the  current oscillations appears as Landau-Zener tunneling transitions  between tilted Landau levels. The hopping distance $\Delta Y$  for the resonant transition  given in Eq. (\ref{eq:hoppdist}) satisfies
 $\Delta Y  = N v_F / \omega_H$, where  $N$ is an integer.
 It is argued that HIRO may be observable in ultra clean graphene with  mobilities above  $\mu \approx 10^{5}  cm^{2} V^{-1} s{-1} $. Finally we also mention that it will be interesting to look after the possibility of producing 
 zero-differential resistance states, similar to those observed in 2DEG \cite{Yang}.



\section*{Appendix: COVARIANT MIGRATION CENTER THEORY}

In this appendix,  we work out the details of the covariant migration center   formalism. 
First it is convenient to write the equation of motion in a manifestly covariant  form 
\begin{equation}\label{covDirac}
{\cal H}  \Psi  \equiv  \left(  v_F \gamma^\mu \Pi_\mu  + \gamma^0 V \right) \Psi = 0  \,, 
\end{equation}
where we recall that the Dirac matrices  are  select as  $\gamma_{\nu} = \left(\sigma_{z} , i \sigma_{y} , -i\sigma_{x} \right)$\cite{Sharapov}. The equation applies both in the $S^\prime_L $ frame (Eq. \ref{eqDirac}), as well  in the boosted frame $S_A$ (Eq. \ref{eqDirac2}). To implement the Schwinger proper time method, suppose we seek for the Green function 
$G (x, x^\prime) $ solution   of $ {\cal H} (x)  G (x, x^\prime) = \delta^{(3)} (x,x^\prime) $. Hence on considers $ {\cal H} (x)$  as a Hamiltonian that describes the proper time $\tau$ evolution of the system
\begin{equation}\label{proptime}
i \hbar \frac{\partial  \Psi}{\partial \tau} =  {\cal H}  \Psi   \,.
\end{equation}
The Green function  $G (x, x^\prime)$ is obtained in terms of the unitary evolution operator $U(x, x^\prime, \tau)$
by means of the relation 
\begin{equation}	
G (x, x^\prime) =  -i \int_{-\infty}^{0} U(x, x^\prime, \tau) d\tau     \,.
\end{equation}
 The  evolution operator satisfies the equation  \begin{equation}	
  i \hbar \frac{ \partial U(x, x^\prime, \tau) }{ \partial \tau } =  {\cal H  } (x) U(x, x^\prime, \tau)   \,.
\end{equation}
with the following boundary conditions: $ lim_{\tau \to 0} \,\,  U(x, x^\prime, \tau) =  \delta^{(3)} (x,x^\prime)$ and 
$ lim_{\tau \to -\infty} \,\,  U(x, x^\prime, \tau) = 0 $.
For the purpose of this paper, we are mainly interested in deducing the covariant guiding center velocities. First we consider the   Heisenberg equation of motion for the $x^\mu $ coordinates, it gives
\begin{equation}	\label{eq:covprob}
\frac{d x^\mu}{d \tau} =\frac{i}{\hbar}\left[ {\cal H },  x^\mu \right]=   v_{F}  \gamma^\mu  \,.
\end{equation}
The Dirac probability current,
is then written as  $j^{\mu} = \bar{\psi}   \gamma^\mu \psi =  \psi^{\dagger}  \gamma^0 \gamma^\mu \psi$.
It is easily verified that the probability current transform as a covariant $2+1$-vector under the Lorentz-like transformations Eqs. (\ref{boost1},\ref{boost2}):  $\tilde{j}^{\mu} =   \Lambda^{\mu}_{\nu} j^{\nu}$.
As explained in section \ref{curr} the electron  coordinate $x^\mu $ splits into the guiding center $X^\mu$ and the   cyclotron or   relative coordinate  $  \eta^\mu$, $i.e.$ $ x^\mu = X^\mu +   \eta^\mu$, where a manifestly covariant expression  for the relative coordinate   $ \eta_\mu $  is given in terms of the velocity operator $\Pi_\mu$ and the 
the electromagnetic tensor $F_{\mu \nu}$ in   Eq. (\ref{eq:covrelative});  where
$F_{\mu \nu}$ is   constructed from the constant fields: $\frac{c}{ v_F}\textbf{E}$ and  $\textbf{B}$ as
\begin{equation}\label{fmunu}
 F_{\mu \nu} = \left( \begin{array}{ccc}
 0 & \frac{c}{v_F} E_x & \frac{c}{v_F} E_y   \\
-\frac{c}{v_F} E_x &  0   & -B \\
-\frac{c}{v_F} E_y & B & 0   \\
\end{array} \right)  \, \end{equation}
The quantity $F_{\mu \nu} \, F^{\mu \nu} = B^2 - \frac{c^2}{v_F^2} E^2  $ is an invariant under the Lorentz transformation.
In the  $S^\prime_L$ frame the three-vector $\eta^{\mu}$ is  given by the expression in Eq. (\ref{eq:relativeSL}), whereas in the  $S_A$ frame it  is  given by usual result  $ \tilde{ \eta}^{\mu}   =  (c/e \tilde{B})(0, -\tilde{ \Pi}_y, \tilde{\Pi}_x)$.
It can be readily  demonstrated that $ \tilde{ \eta}^{\mu}$  transform as covariant three-vector under the Lorentz  transformations Eqs. (\ref{boost1}):  $\tilde{\eta}_{\mu} =   \Lambda_{\mu}^{\nu}  \eta_{\nu}$.
With the aid or the Heisenberg equations of motion and using  the definition for $ \tilde{ \eta}^{\mu}$,  
 the relative proper velocity $  d \eta^\mu / d \tau $ can be obtained both in the $S_L^\prime$ and in the $S_A$ frames.  In the boosted  $S_A$ frame we obtain:
\begin{equation}	
\frac{d \tilde{\eta} ^\mu}{d \tau} = \left(0, v_F \,   \boldsymbol{ \gamma} - \frac{c}{e \tilde{B} } \gamma^0\textbf{e}_{3}\times \boldsymbol{ \nabla}  \tilde{V} \right) \,,
\end{equation}
where $\textbf{e}_{3}$ is an unitary vector in the direction of the magnetic field. On the other hand,  $  d \eta^\mu / d \tau $ evaluated  in the $S_L^\prime$ sheds the following result
\begin{equation}	
\frac{d \eta  ^\mu}{d \tau} = \left(- \beta \frac{d \eta  ^1}{d \tau} , \gamma^2 \left[  v_F \,   \boldsymbol{ \gamma} - \frac{c}{e     B } \gamma^0\textbf{e}_{3}\times \left(  \boldsymbol{ \nabla}  \tilde{V}  - \textbf{E} \right)\right] \right) \,,
\end{equation}
notice that the temporal component $ d \eta  ^0 / d \tau$ is obtained in terms of spatial $x$-component. It is a lengthly but straightforward exercise to demonstrate that the probability current defined in terms of the  relative  velocity 
$  \bar{\psi} \left(  d \eta^\mu / d \tau \right) \psi  $ transform as a covariant three-vector. 
Summarizing all these results, and taking into account that the relative coordinate is obtained as $X^\mu = x^\mu - \eta^\mu$,  it immediately follows that   the current density probability  $\cal{J}_\mu $ defined in terms of the guiding center  velocity as 
 $  { \cal J}_\mu  = \Psi^\dagger \gamma_0 \left(  d  X_\mu / d \tau \right) \Psi$, 
is also   a  Lorentz three-vector.   The   guiding center velocity $ d \tilde {X}_\mu / d \tau $  in the $S_A$ frame as given in Eq. (\ref{3velocity})  is used in the calculation of the impurity induced current Eq. (\ref{corriente}).


\end{document}